\documentclass[pre]{revtex4}

\usepackage{graphicx}
\usepackage{dcolumn}
\usepackage{bm}
\usepackage{amssymb}
\usepackage{amsmath}
\usepackage{amsthm}
\usepackage{verbatim}
\usepackage{xcolor}

\newcommand{\Null}[1]{\mathcal{N}(#1)}
\renewcommand{\ker}[1]{\Null}

\newcommand{\vu}{{\bf{u}}}

\begin{document}

\preprint{APS/123-QED}

\title{Self-sustaining oscillations of a falling sphere through Johnson-Segalman fluids}
\author{Young-Ju Lee$^1$ and Chen-Song Zhang$^2$}
\affiliation{$^1$Rutgers University, The State University of New Jersey, Piscataway, NJ, Math Department \\
$^2$NCMIS \& LSEC, Academy of Mathematics and System Sciences, Beijing, China}

\date{\today}

\begin{abstract}
We confirm numerically that the Johnson-Segalman model is able to reproduce the {\it{continual}} oscillations of the falling sphere observed in some viscoelastic models, \cite{Mollinger;Cornelissen;Brule1999,Jayaraman.A;Belmonte.A2002}. The empirical choice of parameters used in the Johnson-Segalman model is from the ones that show the non-monotone stress-strain relation of the steady shear flows of the model. The carefully chosen parameters yield continual, self-sustaining, (ir)regular and periodic oscillations of the speed for the falling sphere through the Johnson-Segalman fluids. Furthermore, our simulations reproduce the phenomena characterized by Mollinger et al.~\cite{Mollinger;Cornelissen;Brule1999}: the falling sphere settles slower and slower until a certain point at which the sphere suddenly accelerates and this pattern is repeated continually.   
\end{abstract}

\pacs{Valid PACS appear here}

\maketitle

\section{Introduction}

Complex fluids can afford a variety of new phenomena, among others, the flow instabilities due to the internal structures that induce nontrivial interactions between fluids and macromolecules therein~\cite{JDGoddard_2003,Bird.R;Curtiss.C;Armstrong.R1987}. Typical flow instabilities observed in viscoelastic flows can be categorized into two distinguishing types: elastic instability and 
material instability \cite{JDGoddard_2003}. Whereas the elastic instability in the polymer fluids is due to the strong nonlinear mechanical properties of the polymer solutions \cite{Groisman;Steinberg2001}, the material instability is believed to have its origin in non-monotone dependence of stress on strain rate~\cite{JDGoddard_2003}. Examples of material instability, such as the shear banding~\cite{ORadulescu_PDOlmsted_2000,CYDLu_PDOlmsted_RCBall_2000}, are observed in wormlike micellar fluids, that are fluids made up of elongated and semiflexible aggregates resulting from the self-assembly of surfactant molecules \cite{Berretbook2006}. 

A recent and striking evidence of the flow instability in the wormlike micellar fluids is discovered by Jayaraman et al.,~\cite{Jayaraman.A;Belmonte.A2002}: a sphere falling in a wormlike micellar fluid undergoes {\textsf{continual}} oscillations without reaching a terminal velocity; see also~\cite{Chen;Rothstein2004}. The {\it{continual}} oscillations of a falling sphere in a wormlike micellar fluid is quite exotic and in contrast to the case of generic polymeric fluids~\cite{MTArigo_GHMcKinley_1998,CBodart_MJCrochet_1994,RAJAGOPALAN_ARIGO_MCKINLEY_1995, Rajagopalan;Arigo;McKinley1996,GHMcKinley_2001}. Note that the oscillation of a falling sphere has been observed as well in another systems by Mollinger et al.,~\cite{Mollinger;Cornelissen;Brule1999} and was stated as an {\it{unexpected}} phenomenon. 
The steady shear flow of the wormlike micellar fluids, in which the falling sphere continually oscillates, is observed to display a  flat region~\cite[FIG.~2(a)]{Jayaraman.A;Belmonte.A2002} in stress-strain relation, which is known to be linked with the non-monotone shear stress-strain rate relations. 

The non-monotone stress-strain relations have been extensively investigated for such instabilities as shear banding  \cite{CYDLu_PDOlmsted_RCBall_2000,FieldingOlmsted2006}, shark-skin and spurt~\cite{CainDenn1988,YRenardy_1995} and the Johnson-Segalman (\textsf{JS}) model \cite{Johnson.M;Segalman.D1977} showing such nonmonotone stress dependence on the strain has successfully provided qualitative agreements with experimental results in these instances \cite{FieldingOlmsted2006}. It is then conjectured in \cite{Jayaraman.A;Belmonte.A2002} that ``the {\it{continual}} oscillations of the falling sphere could be due to the same instability relevant to the non-monotone stress-strain relations" and suggested the \textsf{JS} model may produce the falling sphere experimental results in the wormlike micellar fluids. It is, however, apparent that the velocity fields of a falling sphere in wormlike micelles are much more complicated than the steady shear flows; the fluids are both elongated and sheared in the wake and near the sphere surface. 
\begin{figure}[htp] 
\centering
\includegraphics[width = 0.45\linewidth,height=0.35\linewidth]{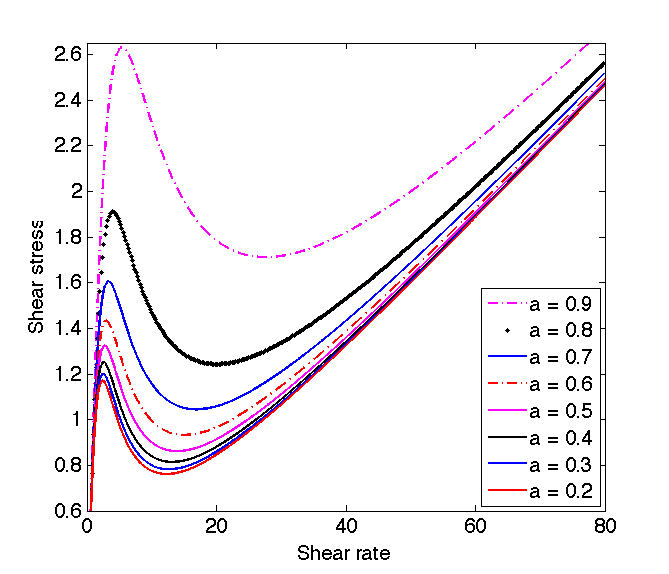}
\includegraphics[width = 0.45\linewidth,height=0.35\linewidth]{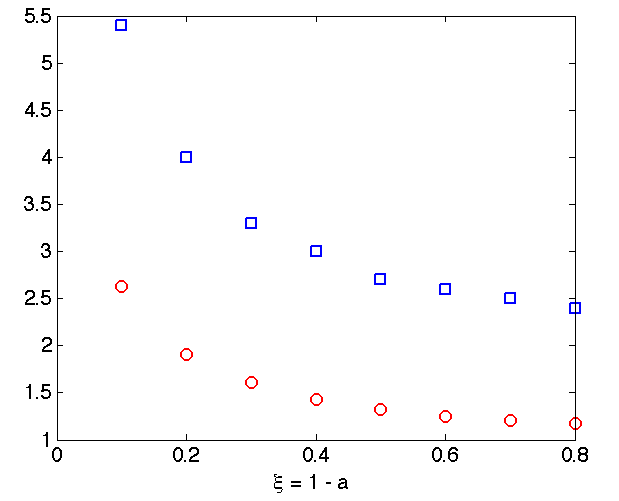}
\vskip -0.3cm
\caption{\footnotesize (Left) The plot of the shear stress as a function of the strain rate for different slippage parameter $\xi = 1 - \textsf{a}$. (Right) The plot of the shear rate at which the local maximum of the shear stress occurs (square) and the local maximum of the shear stress (circle) as a function of the {\it{slippage}} parameter $\xi$. Parameters used are ${\textsf{Re}} = 0$, 
$\mu_s = 0.03$, ${\textsf{Wi}} = 0.45$ and $\xi = 0.1 \sim 0.8$.}\label{slip} 
\end{figure}

In this paper, through a sophisticated numerical technique, we demonstrate that the minimal ingredient that contributes to continual oscillations is related to the non-monotone stress dependence on the strain rate. Moreover, we provide numerical evidences that the flow instability observed in falling sphere experiment~\cite{Jayaraman.A;Belmonte.A2002} can be attributed to the material instability manifested by such a plateau region displayed in the stress-strain rate curve. The rest of the paper is organized as follows. The model description and a stable numerical scheme are presented in Section~\ref{sec:model}.  A number of numerical experiments as well as their physical relevance are presented in Section~\ref{sec:numerics}. Finally, some concluding remarks are given in Section~\ref{sec:conclusions}.  

\section{Flow Models and Numerical Schemes}\label{sec:model}

We assume that a solid sphere of radius $r_s$ with density $\rho_s$ is accelerating with speed $U_s$ from rest under the influence of gravity $g$ in a viscoelastic fluid of density $\rho_f$ with zero shear viscosity $\eta$ contained in a cylinder of radius $r_c$. Let $\wp=\rho_s/\rho_f$ be the density ratio, the symbols $D_t$ and $d_t$ denote the material time derivative and the time derivative, respectively. And, $\bm{e}_z$ is the unit axial vector. The dimensionless momentum equations can be given as follows: 
\begin{equation}\label{eq:cont}
\left\{
\begin{array}{rcl}
{\textsf{Re}} D_t \vu &=& -\nabla p + \nabla \cdot \bm{\tau} + {\textsf{Re}} (d_tU) \, \bm{e}_z
\\
\nabla \cdot \vu &=& 0,
\end{array}
\right.
\end{equation}
where $U$ is the dimensionless speed of falling sphere, the parameters $\textsf{Re} = \rho_f U_N r_s/\eta$ and $\mu_s =  \eta_s/\eta$ and the total stress $\bm{\tau} = \mu_s (\nabla \vu + \nabla \vu^T) + \bm{\tau}_p$. For characteristic scales, we used the following characteristic scales~\cite{Rajagopalan;Arigo;McKinley1996}: 
(1) \textsf{length scale}: radius of the sphere $r_s$, 
(2) \textsf{velocity scale}: Stokes terminal velocity 
$U_N = 2r_s^2(\rho_s - \rho_f) g/9 \eta K$ 
with $\eta = \eta_s + \eta_p$, 
(3) \textsf{time scale}: $r_s/U_N$, 
and 
(4) \textsf{stress and pressure scale}: $\eta U_N / r_s$. 

The dimensionless \textsf{JS} constitutive equation for the elastic stress $\bm{\tau}_p$ reads 
\begin{equation}
\bm{\tau}_p + {\textsf{Wi}} \,\delta_E \bm{\tau}_p = \mu_p (\nabla \vu + \nabla \vu^T), \label{M3}
\end{equation}
where ${\textsf{Wi}} = \lambda U_N/r_s$, $\mu_p = \eta_p/\eta$, and the Gordon-Schowalter convected time derivative~\cite{Bird.R;Curtiss.C;Armstrong.R1987} is defined as
\begin{equation}\label{GS}
\delta_E \bm{\tau}_p := D_t \bm{\tau}_p - \mathcal{R}(\vu) \bm{\tau}_p -  \bm{\tau}_p \mathcal{R}(\vu)^T
\end{equation}
with 
$
\mathcal{R}(\vu) = \frac{1}{2}((\textsf{a}+1) \nabla \vu + (\textsf{a}-1) \nabla \vu^T).
$
\begin{figure}[htp]  
\centering
\includegraphics[height=0.35\linewidth]{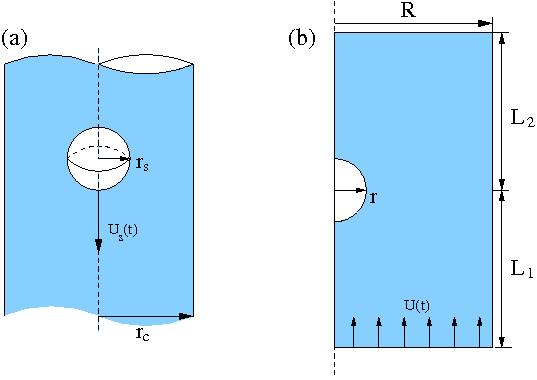}
\caption{\footnotesize A schematic of the domain of the cylinder of radius $r_c$ in which the solid sphere of radius $r_s$ falls with the speed $U(t)$ at time $t$.} 
\label{DOMAIN}
\end{figure}
The parameter $\textsf{a}$ is related to the slippage parameter $\xi = 1 - \textsf{a} \in [0,1]$~\cite{RGLarson_1988}. Note that, when $\xi = 0$, the \textsf{JS} model reduces to the Oldroyd-B (\textsf{OB}) model \cite{JGOLDROYD_1950}.
\begin{figure}[htp] 
\centering
\includegraphics[width = 0.25\linewidth,height=0.25\linewidth]{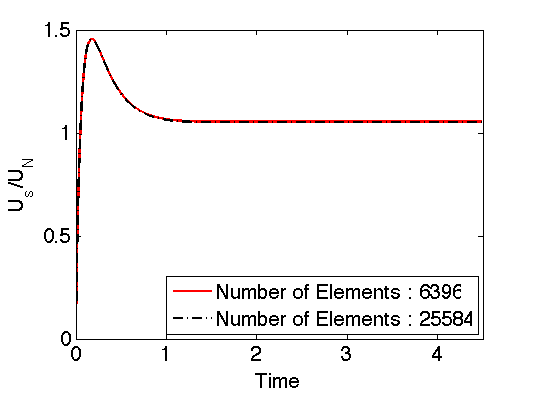}
\includegraphics[width = 0.7\linewidth,height=0.35\linewidth]{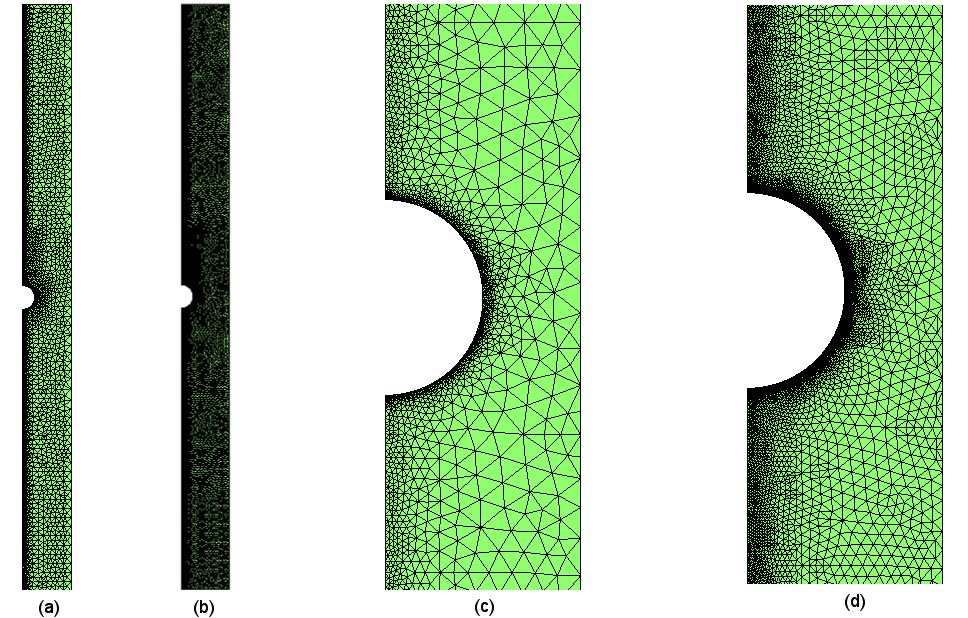}
\caption{\footnotesize Speed of the falling sphere through the Johnson-Segalman models in two different meshes (the fine mesh is refined from the coarse mesh by the regular refinement). 
(Left) Parameters used are \textsf{Re} = 0.0325, $\mu_s = 0.59$, \textsf{Wi} = 0.5, $\alpha = 6.3$ and \textsf{a} = 0.8 and the difference between terminal velocities is 0.002538 (Right) (a) The coarse mesh, \textsf{Mesh I} for the computational domain with the aspect ratio $\wp = 4.115$ in Figure \ref{DOMAIN}. The fine mesh, \textsf{Mesh II}  obtained from \textsf{Mesh I}. (c) Close up view of the mesh for domain with the aspect ratio $\wp = 2$ near the sphere. (d) Close up view of the mesh for domain with the aspect ratio $\wp = 2$ of \textsf{Mesh II} near the sphere.}
\label{fig:nonconv}
\end{figure}

The equation of motion (\ref{eq:cont}) is formulated in the moving frame viewed from the sphere~\cite{Rajagopalan;Arigo;McKinley1996} and is coupled with the dynamic force balance equation of the falling sphere, whose dimensionless form is given by $(2 \textsf{Re} \wp/3) d_t U =
3 K + F_d$ where $F_d$ is the drag force~\cite{JHAPPEL_HBRENNER_1973} and $K$ is the wall correction factor. The constitutive equation (\ref{M3}) can be reformulated in terms of the conformation tensor $\bm{c}$ \cite{Linliuzhang2005}, the ensemble average of the dyadic product of the end-to-end vector for the dumbbells as 
$
\bm{c} + \textsf{Wi} \delta_E \bm{c} = (\mu_p/\textsf{a} \textsf{Wi}) \bm{\delta}, 
$
where $\bm{\delta}$ is the identity tensor. The conformation tensor $\bm{c}$ of {\textsf{JS}} model can be shown to be positive definite at all time~\cite{Lee;Xua2006,MAHULSEN_1990,Beris.A;Edwards.B1994}. 
In our numerical simulations, we assume that the flow is axisymmetric and apply the numerical methods proposed in~\cite{Lee;Xua2006}, which is based on the semi-Lagrangian method (\textsf{SLM})~\cite{Pironneau1982,Douglas1982} and preserves the positivity of the conformation tensor in the discrete level. Let $(\vu^{\rm old},p^{\rm old},\bm{c}^{\rm old})$ denote the previous semi-discrete solutions. The backward Euler method combined with the \textsf{SLM} method applied for both $D_t \vu$ and $D_t \bm{c}$ result in the following semi-discrete systems which will define the current time level solutions $(\vu^{\rm new},p^{\rm new},\bm{c}^{\rm new})$. 
\begin{equation} 
({\textsf{Re}}/h_t) \vu^{\rm new} + \nabla p^{\rm new} - \mu_s \Delta \vu^{\rm new} 
= (\textsf{Re}/h_t) \vu^{\rm old}\circ y + \nabla \cdot \bm{c}^{\rm old} + {\textsf{Re}} (d_t U)^{\rm old} \,\bm{e}_z,
\end{equation}
and $\bm{A}\bm{c}^{\rm new} +\bm{c}^{\rm new}\bm{A}^T = \bm{C},$ with $
\bm{A} = 1/2 + \textsf{Wi} h_t/2- \mathcal R(\vu^{\rm new})$
and $\bm{C} = (\mu_p/\textsf{a} {\textsf{Wi}}) \bm{\delta} + (\textsf{Wi}/h_t) \bm{c}^{\rm old} \circ y$,
where $y$ denote the characteristic feet of the particle $X$ at the current time, which can be obtained by solving the flow map equation that $d_s y(X,s) = \vu(y(X,s),s)$ and $y(X,t) = X$, $h_t$ is the time step size, $(d_t U)^{\rm old}$ is the acceleration of the falling sphere at $t^{\rm old}$. 

Note that the semi-discrete constitutive equation for the constitutive equation can be viewed as the Lyapunov equation~\cite{Lancaster;Rodman;1980}. For the spatial discretization, we apply the standard finite element approximations: continuous piecewise quadratic polynomial approximation for the velocity field, continuous piecewise linear approximation for the pressure, and continuous piecewise linear polynomial approximation for the conformation tensor. More detailed description of our algorithms can also be found at the review paper \cite{booklee2011}. Note that the choice of our finite elements for the velocity and pressure has been shown to be stable for the axisymmetric Stokes equations \cite{lee;li2011} recently.  
For all simulations, we use the time step size $h_t = 10^{-3}$. The numerical solutions have been shown to converge with respect to the mesh size in case they achieve the steady-state; FIG.~\ref{fig:nonconv} (Left) indicates this visibly and the mesh convergence is observed.

\section{Numerical Experiments and Discussions}\label{sec:numerics}

\begin{figure}[htp] 
\centering
\includegraphics[width = 0.8\linewidth,height=0.35\linewidth]{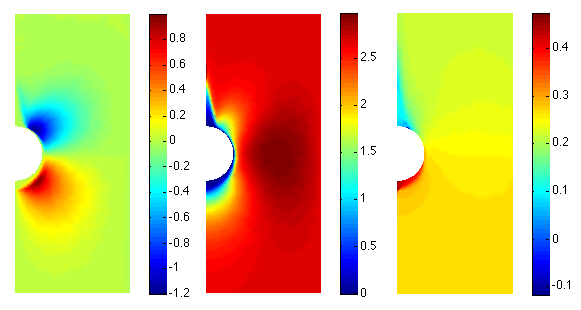}
\caption{\footnotesize The contour plot of velocity fields u (Left), v (Middle) and pressure, p (Right) from the computation of the Johnson-Segalman model at an instant time level, $t = 25.665$. The parameters are \textsf{Re} = 0.0325, $\mu_s = 0.03$, \textsf{Wi} = 0.45 and the density ratio $\alpha = 6.3$.}
\label{fig:velcon}
\end{figure}

\begin{figure}[htp] 
\centering
\includegraphics[width = 0.9\linewidth,height=0.35\linewidth]{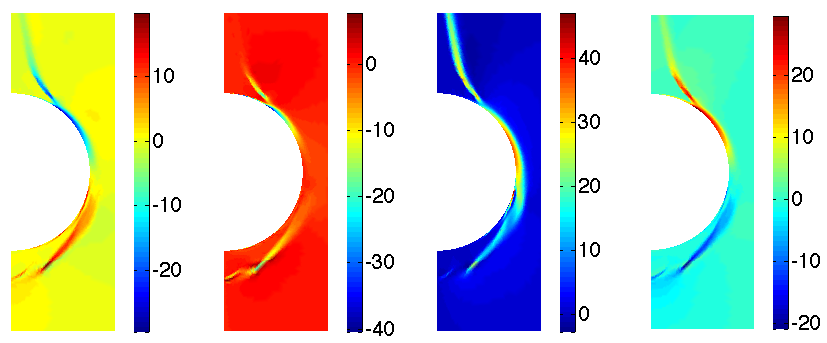}
\vskip -0.4cm
\caption{\footnotesize The contour plot of gradient of velocity fields $\partial_r u, \partial_z u, \partial_r v$ and $\partial_z v$ viewed near from sphere from left to right, respectively of the Johnson-Segalman fluid at time level $t = 8.5367$. Here $(u,v)$ are the radial and axial velocity component of $\vu$. The parameters are \textsf{Re} = 0.0325, $\mu_s = 0.03$, \textsf{Wi} = 0.45, $\wp = 6.3$ and $\alpha = 4.115$}
\label{fig:velcon}
\end{figure}

For the numerical experiments, we focus on the choice of slippage parameter denoted by $\xi$. In particular, we have chosen parameter ranges that show the non-monotone stress-strain rate for the one dimensional parallel shear flows, for which the shear stress and strain rate $\kappa$ relation is
$\tau(\kappa) = \mu_s \kappa + (\zeta \mu \kappa)/[9\zeta^2 + \xi ( 2- \xi) \kappa^2]$, where $\zeta = 1/{\textsf{Wi}}$ and $\mu = (1-\mu_s)/{\textsf{Wi}}$. We note that the non-monotone stress-strain relation can always be obtained for $\xi >0$; see Fig.~\ref{slip} (Left). It is observed that not all such parameters yield continual oscillations. For example, while $\xi = 0.2$ gives the non-monotone shear stress-strain relation, it does not yield {\it{continual}} oscillations, which indicates that the choice of parameters based on the simple shear flows can be misleading. Our results, however, demonstrate these curves are apparently related to the observed flow instability in the wormlike micellar fluids. More precisely, by varying the parameter $\xi = 0.0 \sim 0.8$, we observe that the larger $\xi$ is, the smaller the local maximum value of the shear stress at which the transition to the negative slope of the shear stress curve occurs; see FIG.~\ref{slip}. 

\begin{figure}[htp] 
\centering
\includegraphics[width = 0.45\linewidth,height=0.35\linewidth]{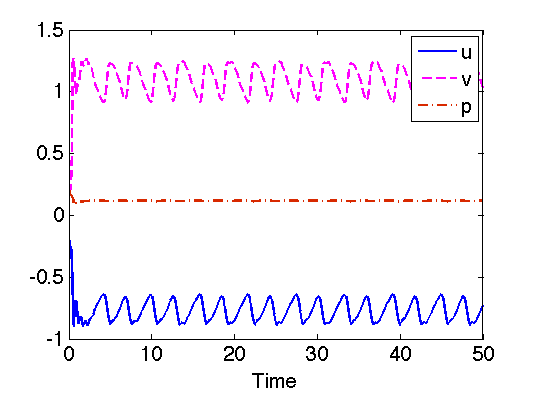}
\includegraphics[width = 0.45\linewidth,height=0.35\linewidth]{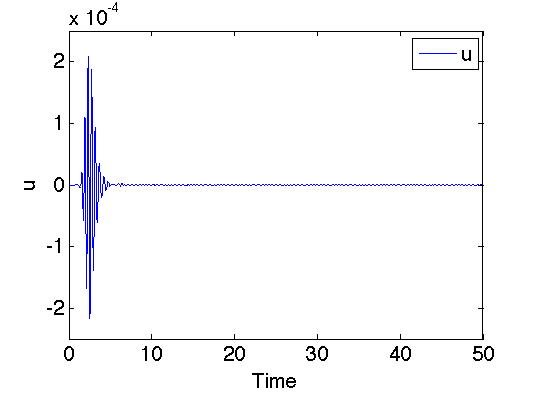}
\vskip -0.4cm
\caption{\footnotesize (Left) The velocity fields $u$ (solid), $v$ (dot) and pressure $p$ (solid dot) as a function of time at specified point $\bm{x}_1$ in the domain. (Right) The plot of $u-$component of the velocity $\vu$ at the point $\bm{x}_2$. The parameters used are : {\textsf{Re}} = 0.0325, $\mu_s = 0.03$, ${\textsf{Wi}} = 0.45$, $\textsf{a} = 0.3$ and the aspect ratio is $\alpha = 4.115$.}
\label{fig:fieldosc}
\end{figure}

The range $\xi = 0.4 \sim 0.8$ produces continual oscillations and the \textsf{amplitude} of the oscillations is a {\it{increasing}} function w.r.t. $\xi$; see FIG.~\ref{fig:amplitude}. The parallel shear flows have been studied by Renardy \cite{YRenardy_1995}, which indicate that the apparent local shear banding observed in FIG.~\ref{fig:velcon} can in fact, be related to the flow instability. Unlike one dimensional shear flows \cite{Nohel;Pego1993,GCGeorgiou_DVlassopoulos_1998}, it has been shown that the shear flows is of the short-wave instability and unstable with respect to the two dimensional perturbation. Our numerical solutions further support important aspects of theoretical results on the material instability by Renardy~\cite{YRenardy_1995}, namely, the locality of the material instability. The instability in the parallel shear flow of \textsf{JS} model is shown to possess the feature: ``the most unstable mode is found at the interfacial mode and the short-wave instability decays exponentially fast away from the interface position and the effect is localized in a boundary layer near the interface, not disturbing the bulk of the flow". We consider two points: $\bm{x}_1$ is located near the sphere, whose distance from the surface of the sphere is $0.2293$ and $\bm{x}_2$ is chosen away from the sphere, near the top wall and compare the solution behavior by plotting sample variables, the velocity fields $\vu =(u,v)$ and the pressure $p$ in the FIG.~\ref{fig:fieldosc}. Whereas at $\bm{x}_1$, all solutions are unsteady, at $\bm{x}_2$ the velocity component $u$ is steady. This fact may justify our assumption on the flow symmetry.
 
\begin{figure}[htp]  
\centering
\includegraphics[width = 0.4\linewidth,height=0.35\linewidth]{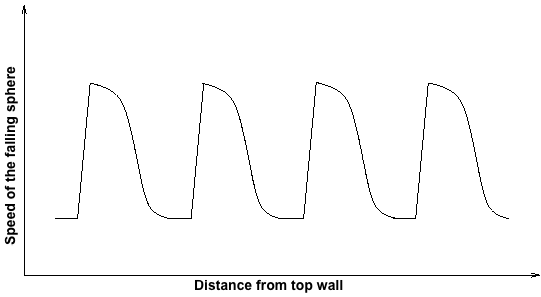}
\includegraphics[width = 0.45\linewidth,height=0.35\linewidth]{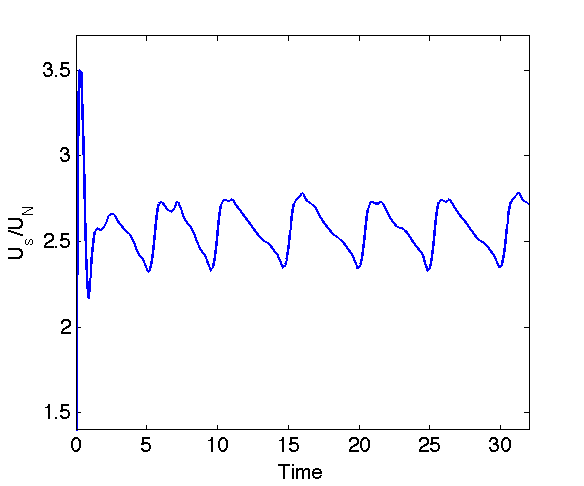}
\vskip -0.3cm
\caption{\footnotesize (Left) The schematic speed of the falling sphere as a function of the distance from the top wall~\cite{Mollinger;Cornelissen;Brule1999}. (Right) Typical oscillatory speed of a sphere falling through \textsf{JS} fluids.}
\label{Oscillsanand}
\end{figure} 
Our numerical results agree qualitatively with the experimental observations, \cite{Mollinger;Cornelissen;Brule1999,Jayaraman.A;Belmonte.A2002} that the large sphere oscillates whereas the small sphere does not oscillate, which indicates that the continual oscillations may be affected by the wall effects. More precisely, we observed that the class of parameters, which produced the continual oscillation of falling sphere for the aspect ratio $\alpha = 4.115$, did not produce the continual oscillations for the aspect ratio $\alpha = 8.115$ while the amplitude of the oscillation for $\alpha = 6.115$ is smaller than that for the case $\alpha = 4.115$. Furthermore, the oscillation pattern is clearly similar to the physical observations stated by Mollinger et al., \cite{Mollinger;Cornelissen;Brule1999}. Namely, the falling sphere settles slower and slower until a certain point at which the particle suddenly accelerates and this pattern is repeated continually, see FIG \ref{Oscillsanand} and FIG \ref{fig:nonconv}. Note that this characteristic can be found at the work by Jayaraman et al., \cite{Jayaraman.A;Belmonte.A2002} as well.

\begin{figure}[htp] 
\centering
\includegraphics[width = 0.9\linewidth,height=0.35\linewidth]{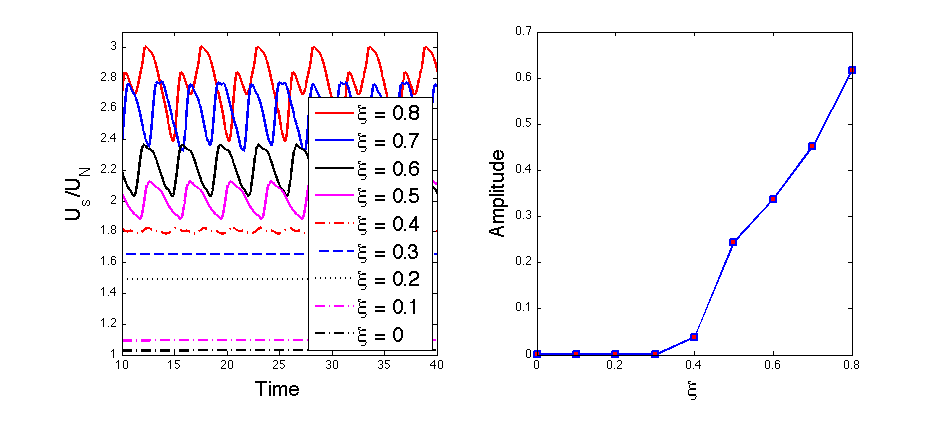}
\vskip -0.5cm
\caption{\footnotesize (Left) The speed of the falling sphere as a function of time for different choices of the slippage parameter $\xi$. (Right) Amplitude of the oscillations as a function of the {\it{slippage}} parameter $\xi$.}\label{fig:amplitude}
\end{figure}
\begin{figure}[htp] 
\centering
\includegraphics[width = 0.45\linewidth,height=0.35\linewidth]{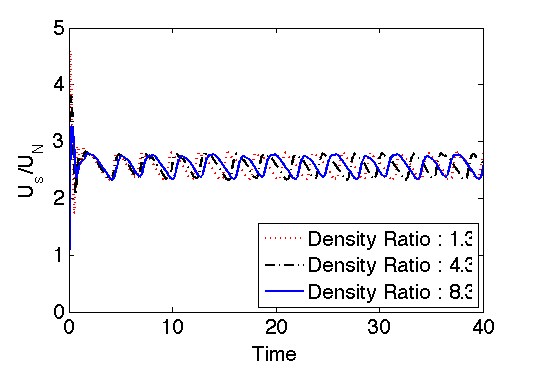}
\includegraphics[width = 0.45\linewidth,height=0.35\linewidth]{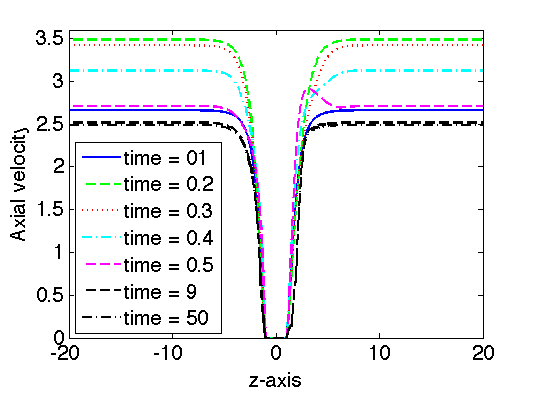}
\vskip -0.4cm
\caption{\footnotesize (Left) Normalized speed of the falling ball in {\textsf{JS}} fluids as a function of time varying the density ratio $\wp$ from 1.3 to 8.3. (Right) The plot of fluid velocity along the cylinder axis for \textsf{JS} model in time evolution with density ratio $\wp = 6.3$. For both computations, the parameters used are : {\textsf{Re}} = 0.0325, $\mu_s = 0.03$, ${\textsf{Wi}} = 0.45$, $\textsf{a} = 0.3$ and the aspect ratio is $\alpha = 4.115$.
}\label{fig:density}  
\end{figure}

The flow past a sphere in the wormlike micellar fluids is observed to generate negative wake repeatedly before the falling sphere accelerates during the oscillations~\cite{Jayaraman.A;Belmonte.A2002}, i.e., the fluid behind the sphere moves in the opposite direction to the falling sphere,~\cite{OHassager_1979,MTArigo_GHMcKinley_1998}. The \textsf{JS} model produced the temporarily developed negative wake which disappears in time, see FIG.~\ref{fig:density}. This observation is similar to what is observed for the time dependent simulations of \textsf{OB} model, \cite{OGHarlen_2002} and it may be not surprising since the formation of the negative wakes are observed for the class of FENE models~\cite{OGHarlen_2002,JVSatrape_MJCrochet_1994} at sufficiently low values of the extensibility parameter $L$ of the dumbbell~\cite{OGHarlen_2002,JVSatrape_MJCrochet_1994} and both \textsf{OB} and \textsf{JS} models do not impose the finite extensibility of the dumbbells. This result indicates that the negative wake does not appear to be closely related to the {\it{continual}} oscillations of the falling sphere. Finally, unlike what is observed in physical experiments, \cite{Jayaraman.A;Belmonte.A2002} that oscillations can occur for larger density ratios between the sphere and the fluids; while a Delrin sphere of density $\rho_s = 1.35 g/cm^3$ with the diameter $d = 3/16\,{\rm in}$ does not oscillate, a Teflon sphere of density $\rho_s = 2.17 g/cm^3$ of the same diameter does oscillate. As demonstrated in FIG.~\ref{fig:density}, the results shows that the simulation of {\textsf{JS}} model with three different density ratios, $\alpha = 1.3, 4.3$ and $8.3$ produce similar pattern, which provide a room yet to be improved in mathematical modeling of the wormlike micellar fluids.

\section{Conclusions}\label{sec:conclusions}
In conclusions, the \textsf{JS} model is shown to reproduce the continual oscillations at certain range of parameters showing the non-monotone stress-strain rate relations, which is, therefore, believed to be the minimal ingredient to trigger the flow instability relevant to {\it{continual}} oscillations. Some qualitative pattern of the {\it{continual}} oscillation of the falling sphere has been observed as well. For more quantitative agreements with the real experiments, one has to employ the fundamental modeling and simulations of the microstructures of the wormlike micelles such as breaking and reforming of the wormlike micelles. This will be the worthy challenges in the future research. At this point, we are performing our simulation in the full three dimensional setting in order to eliminate the assumption that the flow is symmetric. 
 
\bigskip 
\noindent{\bf Acknowledgement} The authors acknowledge Professors Jinchao Xu, Chun Liu, Andrew Belmonte, Michael Renardy, Yuriko Renardy, and Drs.~Anand Jayaraman, Kensuke Yokoi for helpful discussions. The authors also thank the High Performance Computing Group at the Penn State University to allow the simulation to be done on their cluster. Lee thanks the hospitality of the IMA (Institute of Mathematics and Applications) where this paper has been completed. Lee is partially supported by NSF  DMS-0915028 and Start-up funds from Rutgers University. Zhang is partially supported by Dean Startup Fund, Academy of Mathematics and System Sciences and NSFC 91130011. 
 
\bibliography{rutgerslee}

\end{document}